\documentclass[slac_one]{revtex4}
\usepackage{graphicx}
\usepackage{fancyhdr}
\begin{document}

\title{Open heavy flavour reconstruction in the ALICE central barrel} 

\author{F. Prino for the ALICE collaboration}
\affiliation{INFN, Sezione di Torino}

\begin{abstract}
The ALICE experiment will be able to detect open charm and beauty 
hadrons in proton-proton and heavy ion collisions in the new energy regime 
of the CERN Large Hadron Collider (LHC).
Heavy flavours are a powerful tool to investigate the 
medium created in high energy nucleus--nucleus 
interactions because they are produced in the hard scatterings occurring
at early times and, thanks to their long lifetime on the collision 
timescale, they probe all the stages of the system evolution.
The detectors of the ALICE central barrel ($-0.9 < \eta < 0.9$) will allow
to track charged particles down to low transverse momentum 
($\approx$ 100 MeV/$c$) and will provide 
hadron and electron identification as well as an accurate measurement of 
the positions of primary and secondary vertices.
It will therefore be possible to measure the production of open heavy flavours 
in the central rapidity region down to low transverse momentum, exploiting 
the semi-electronic and the hadronic decay channels.
Here we present a general overview of the ALICE perspectives 
for heavy flavour physics and some examples from the 
open charm and beauty analyses which have been developed and tested on 
detailed simulations of the experimental apparatus.
\end{abstract}

\maketitle

\thispagestyle{fancy}

\section{INTRODUCTION} 

The measurement of the production of open charm and beauty hadrons is a 
powerful tool to investigate the properties of the dense and hot medium 
created in ultra-relativistic heavy ion collisions.
Due to the large mass of $c$ and $b$ quarks, the production of $c\bar{c}$ 
and $b\bar{b}$ pairs can only occur in primary hard scatterings with large 
virtualities.
Hence, the open heavy flavour production cross section in nucleon-nucleon
collisions can be calculated in the framework of the factorization
theorem starting from the DGLAP evoluted 
Parton Distribution Functions (PDF) and Fragmentation Functions and 
from the heavy quark production cross section at the partonic level.
This last element can be calculated with perturbative QCD beyond the
LO~\cite{MNR,FONLL}.
In nucleus-nucleus collision, heavy flavour production can be evaluated 
starting from the nucleon-nucleon pQCD calculations and assuming scaling 
with the number of inelastic nucleon-nucleon collisions (binary scaling).
In this framework, the heavy ion collision is modeled as a superposition
of several independent nucleon-nucleon collisions:
for central Pb-Pb collisions at the LHC ($\sqrt{s}$=5.5 TeV) the number 
of inelastic nucleon-nucleon collisions (N$_{coll}$) will be 
$\approx$1800, leading to abundant heavy flavour production.

Such binary scaling is however broken by the presence of initial and final 
state effects.
Initial state effects are modifications of the PDF inside the nuclei,
parton saturation at small x,
and k$_{\rm T}$ broadening due to Cronin effect.
Final state effects could be due to the presence of the medium created in the 
heavy ion collisions.
A coloured parton is predicted to lose energy while traversing a coloured 
medium 
(with deconfined quarks and gluons) both by radiative and collisional 
mechanisms~\cite{Eloss}.
The experimental observable that is used to study these effects 
is the nuclear modification factor R$_{\rm AA}$ defined as
\begin{equation}
R_{AA}(p_T)=\frac {d^2N_{AA}/dp_T dy}{\langle N_{coll} \rangle d^2N_{pp}/dp_T dy }
\label{eq:RAA}
\end{equation}
which describes the deviation with respect to binary scaling.
Energy loss is expected to be different for quarks and gluons,
thus leading to the expectation of different quenching for
heavy flavoured hadrons (mostly coming from a quark jet) and
light hadrons (mostly coming from gluon jets).
Furthermore, due to their large mass, the energy loss for open beauty hadrons 
is expected to be reduced by the dead cone effect~\cite{DC}.
These features can be studied via the double ratios~\cite{ADSW}:
\begin{equation}
R_{Dh}(p_T)=\frac {R_{AA}^{\rm D~mesons}(p_T)}{R_{AA}^{\rm light~hadrons}(p_T)}
~~~~~~~~~~
R_{BD}(p_T)=\frac {R_{AA}^{\rm B~mesons}(p_T)}{R_{AA}^{\rm D~mesons}(p_T)}
\label{eq:doubleR}
\end{equation}
which will be accessible at the LHC thanks to the abundant production of $c$ 
and $b$ quarks (see next section).

In case of substantial energy loss (as the one observed at RHIC and 
the larger one anticipated for the LHC), heavy quarks may result to be
strongly coupled with the azimuthally asymmetric medium and participate 
in the collective motion (flow) which 
develops as a consequence of the re-scatterings among the 
produced particles.
For non-central collisions, as a consequence of the geometrical 
anisotropy of the overlap region of the colliding nuclei,
the heavy quark thermalization in the early stages of the system evolution
gives rise to an anisotropic flow in the transverse plane.
The development of such a collective motion generates
a sizable value of the Fourier coefficient which 
describes an elliptic azimuthal anisotropy of the observed particles:
$v_2=\langle \cos [ 2 (\varphi- \Psi_{\rm RP})] \rangle$,
where $\Psi_{\rm RP}$ is the reaction plane angle defined by 
the impact parameter vector in the transverse plane.
A contribution to $v_2$ can also result from azimuthal dependent 
energy loss due to the initial geometrical anisotropy of the fireball.

Finally, the quenching due to energy loss may influence also
the hadronization mechanisms at low/intermediate momenta.
In this domain, the hadronization of the slowed down heavy quarks
is expected to occur mainly through quark coalescence in the medium,
thus modifying the relative abundances of particle species with respect to 
the case of hadronization via parton fragmentation in the vacuum.
In particular, this would lead to an increased baryon/meson ratio as 
well as to an enhanced
relative abundance of hadrons containing strange quarks.

\section{CHARM AND BEAUTY AT THE LHC}

At LHC energies ($\sqrt{s}$= 14 TeV for p-p and 
$\sqrt{s_{NN}}$=5.5 TeV for Pb-Pb collisions), charm and beauty 
production will be abundant: the cross section increases by about
a factor 10 for charm and 100 for beauty with respect to RHIC top energy.
The charm and beauty cross-sections used in simulations of p-p collisions 
at $\sqrt{s}$= 14 TeV are obtained from NLO pQCD calculations resulting in
0.16 $c\bar{c}$ and 0.007 $b\bar{b}$ pairs per event~\cite{PPR2}.
For nucleus-nucleus interactions, binary scaling is assumed 
and the nuclear modification of the PDF due to shadowing are taken 
into account, obtaining 115 $c\bar{c}$ and 4.6 $b\bar{b}$ pairs 
per central (0-5\%) Pb-Pb event.
The large yields of $c$ and $b$ quarks will allow detailed studies on the
heavy flavour energy loss as well as on the possible charm and beauty
thermalization in the QCD medium.
On this respect, in order to constrain theoretical models, 
it is of crucial importance to have an experimental apparatus capable
of measuring separately open charm and beauty hadrons.

The large collision energy will allow also to explore 
unprecedentedly small values of Bjorken x.
By reconstructing open charm hadrons at low p$_{\rm T}$, it is possible to
reach x values as small as 10$^{-4}$ at mid-rapidity, thus
opening the possibility of investigating
saturation effects which are expected to play a crucial role in this
x region.

\section{ALICE POTENTIAL FOR OPEN HEAVY FLAVOURS}

The ALICE apparatus~\cite{TPAP} has excellent capabilities for heavy flavour
measurements, for both open heavy flavoured hadrons and quarkonia.
In this paper, we will limit the discussion to the detection of open charm
and beauty in the central barrel and therefore only the detectors
involved in these analyses are described in the following.

The ALICE central barrel covers the pseudo-rapidity region
$-0.9 < \eta < 0.9$ and is equipped with tracking detectors
and particle identification systems embedded in a magnetic field
B=0.5 T.
The combined information from the central barrel detectors allows to track 
charged particles down to low transverse momenta 
(low p$_{\rm T}$ cut-off $\approx$ 100 MeV/c) and provides 
hadron and electron identification as well as an accurate measurement 
of the positions of the primary (interaction) vertex and of
the secondary (decay) vertices.
The main tracking detector is the Time Projection Chamber (TPC) 
which provides track reconstruction and particle identification 
via dE/dx.
The Inner Tracking System (ITS) is the central barrel detector
closer to the beam axis and is composed of six cylindrical layers 
of silicon detectors. 
The two innermost layers (at radii of $\approx$ 4 and 7 cm)
are equipped with pixel detectors, the two intermediate layers
(radii $\approx$15 and 24 cm) are made of drift detectors, while
strip detectors are used for the two outermost layers (radii
$\approx$39 and 44 cm).
The ITS is a key detector for open heavy flavour studies because it 
allows to measure the track impact parameter 
(i.e. the distance of closest approach of the track to the primary vertex)
with a resolution better than 50 $\mu$m for $p_{\rm T}>$ 1.3 GeV/$c$, 
thus providing the capability
to detect the secondary vertices originating from heavy flavour decays.
Two other systems play an important role in the heavy flavour 
analyses as far as particle identification is concerned.
They are the Transition Radiation Detector (TRD) for high-momentum electron
identification and the Time-Of-Flight (TOF) for pion, kaon and proton 
separation on the basis of their time of flight to the TOF.
All these four detectors have full azimuthal coverage.

In the Monte Carlo studies reported here, charm and beauty hadrons have been
generated with PYTHIA~\cite{PYTHIA} tuned to reproduce the heavy quark 
abundances and $p_{\rm T}$ spectra predicted by NLO pQCD~\cite{MNR,PPR2}.
For Pb-Pb collisions, the underlying event has been simulated assuming
dN/dy=6000 in the central rapidity region.
The generated particles are propagated through the ALICE apparatus
using a detailed description of the response of the various detectors.

\subsection{Open charm reconstruction}

The exclusive reconstruction of D mesons from hadronic decays
has been studied with detailed simulations of the ALICE apparatus
for the channels $D^0\rightarrow K^-\pi^+$~\cite{PPR2} and
$D^+\rightarrow K^-\pi^+\pi^+$~\cite{Bruna}.
The reconstruction of other channels, namely
$D^0\rightarrow K^-\pi^+\pi^+\pi^-$, $D_s^+\rightarrow K^-K^+\pi^+$ 
and $\Lambda_c^+\rightarrow K^-\pi^+p$, is also under
investigation.
It is important to extract the total charm cross-section 
from the largest possible number of independent channels 
in order to minimize the systematics.
Furthermore, ratios of abundances of different D mesons, 
like $D_s^+(c\bar{s})/D^+(c\bar{d})$, 
are expected to be sensitive to the different hadronization mechanisms 
at work (e.g. string fragmentation vs. coalescence).

\begin{figure}
\resizebox{0.48\textwidth}{!}{%
\includegraphics*[]{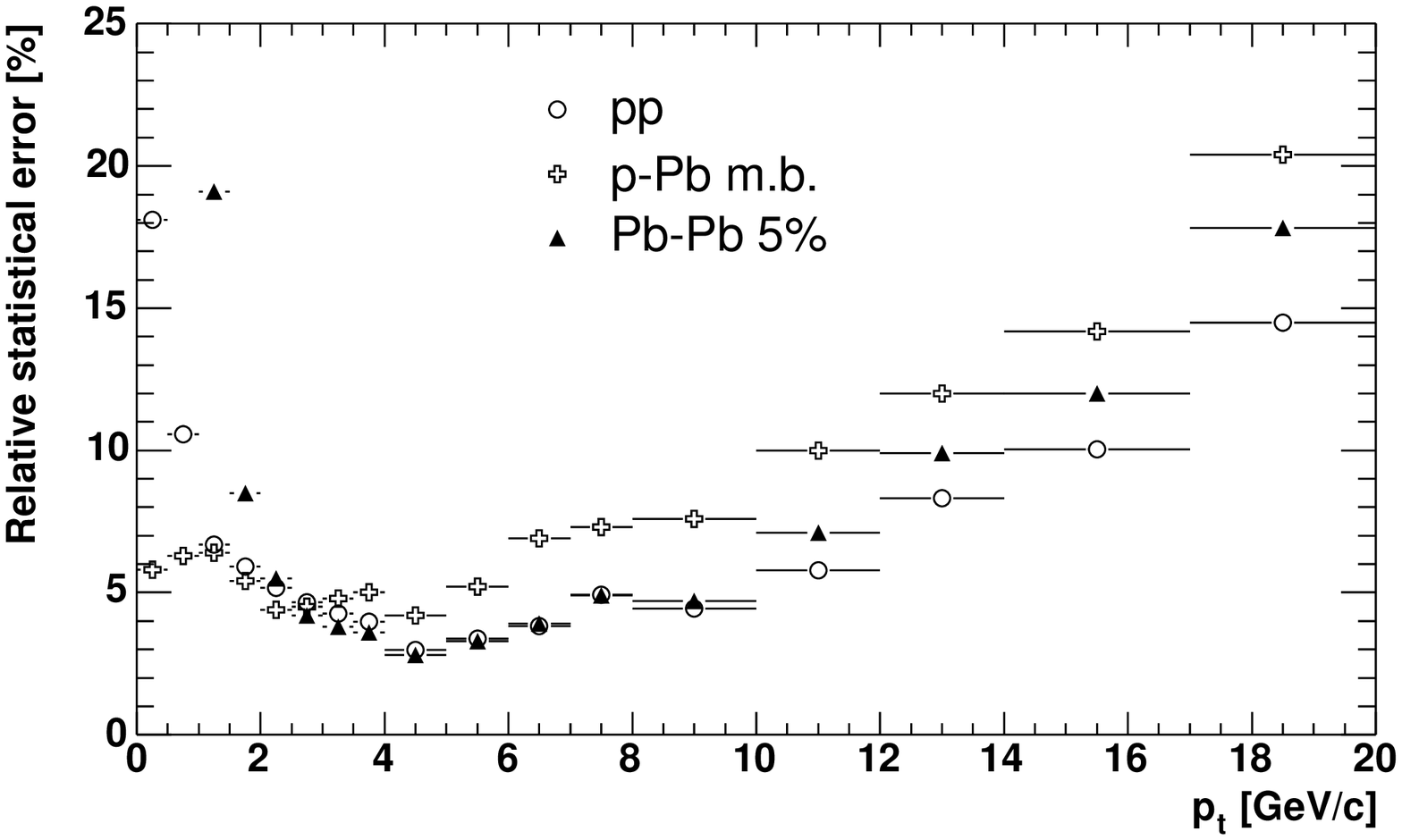}
}
\resizebox{0.46\textwidth}{!}{%
\includegraphics*[]{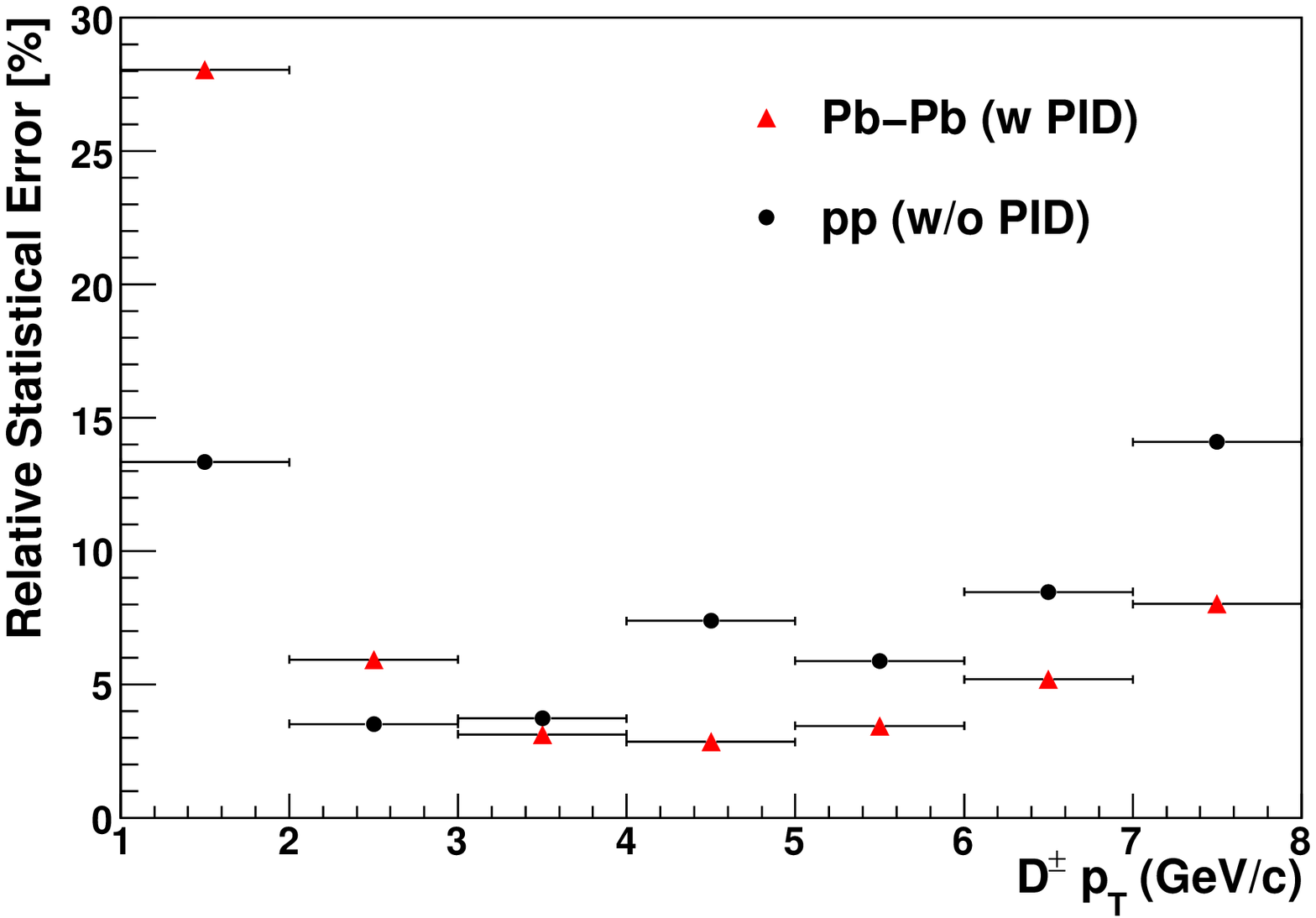}
}
\caption{Expected relative statistical error in 1 year of data taking
for $D^0\rightarrow K^-\pi^+$ (left) and $D^+\rightarrow K^-\pi^+\pi^+$ 
(right).}
\label{fig:staterrc}
\end{figure}

The analysis strategy is based on an invariant mass analysis of
fully reconstructed decay topologies originating from displaced vertices. 
The aim is to identify the tracks coming from open charm 
decays within the full sample of reconstructed tracks.
These tracks are originating from secondary vertices which, 
given the relatively long lifetime of these hadrons 
($\sim$ 0.5-1 ps) are displaced by typically hundreds of microns from the
primary vertex.
Due to the large combinatorial background, selection cuts are quite severe
and have been tuned in order to maximize the statistical significance.
Tracks are first selected according to their transverse momentum and 
their impact parameter.
Then all track combinations with proper charge are considered
and further selection cuts are applied.
In the case of the $D^0$, the main selections are based on the 
product of the impact parameters of the K and $\pi$ candidates 
and on the request that the reconstructed D meson flight line point to 
the primary vertex, i.e. cosine of the pointing angle close to 1.
For the $D^+$, the longer lifetime (c$\tau$= 310 $\mu$m) can be exploited
and the crucial selection variables are the distance between the reconstructed 
primary and secondary vertices and the cosine of the pointing angle as well.
Particle identification information will also be used in the selection, at
least in Pb-Pb collisions.
The relative statistical error (corresponding to the inverse of the
significance) on the $D^0$ and $D^+$ measured yields
expected in 1 year of data taking 
(i.e. 10$^7$ central Pb-Pb and 10$^9$ p-p collisions)
is shown in fig.~\ref{fig:staterrc} as a function of the D meson p$_{\rm T}$.
For the $D^0$ also the case of p-Pb collisions at $\sqrt{s}=8.8$ TeV
is shown.
The results prove the feasibility, with good performance, of exclusive 
$D^0$ and $D^+$ reconstruction in a wide range of transverse momentum 
(1$<$p$_{\rm T}<$20 GeV/$c$).

\subsection{Open beauty reconstruction}

\begin{figure}
\resizebox{0.4\textwidth}{!}{%
\includegraphics*[]{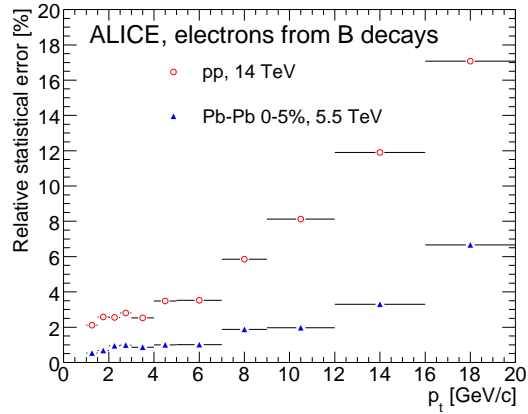}
}
\caption{Expected relative statistical error in 1 year of data taking
for beauty reconstruction from displaced electrons.}
\label{fig:staterrb}
\end{figure}

A detailed Monte Carlo study of inclusive B meson reconstruction from 
semi-electronic decays has been performed~\cite{PPR2}.
The selection strategy exploits the different shapes of the
p$_{\rm T}$ and track impact parameter distributions of the different
electron sources (beauty, charm and the various background contributions).
The crucial quantities for this analysis are the efficiency and the purity of
electron identification in the TPC and in the TRD as well as the resolution 
on the track impact parameter used to select particles displaced from the 
primary vertex.
The contamination from charm semi-electronic decays (which 
is significantly reduced by the impact parameter cut thanks to 
the larger c$\tau$ of B mesons) will be subtracted using the
direct measurement of the charm cross-section as obtained from exclusive
reconstruction of D mesons.
After the selection cuts (in the Monte Carlo studies we have required 
p$_{\rm T}>2$ GeV/$c$ and track impact parameter $>200~\mu$m), 
the expected statistics 
for 10$^7$ central Pb-Pb events is $\approx$ 80000 e$^\pm$
with a purity of about 80\%.
The expected relative statistical error for 1 year of data taking 
(10$^7$ central Pb-Pb and 10$^9$ p-p events) is shown in 
fig.~\ref{fig:staterrb}.

The p$_{\rm T}$ spectra of B mesons down to p$_{\rm T} \approx 0$ GeV/$c$  
can also be obtained from the secondary J/$\psi$ originating
from B decays, as done for instance by CDF~\cite{CDF}.
J/$\psi$ mesons can be detected in the central barrel via their
e$^+$e$^-$ decay exploiting tracking and particle identification in 
ITS, TPC and TRD.
The ITS resolution for the secondary vertices allow to separate
the prompt J/$\psi$ from the ones coming from B decays.
Finally, a promising approach, presently under investigation,
is the topological reconstruction of beauty hadrons 
from unbalanced displaced vertices with large number of prongs~\cite{Faivre}.

\end{document}